\documentclass[amssymb,amsmath,twocolumn,floatfix,aps,showpacs]{revtex4}

\usepackage{graphicx}
\usepackage{dcolumn}
\usepackage{bm}
\usepackage{hyperref,color}

\begin{document}

\title{Efficiency of Single Particle Engines}

\author{Karel Proesmans}
 \email{Karel.Proesmans@uhasselt.be}
 \altaffiliation{Hasselt University, B-3590 Diepenbeek, Belgium.}
\author{Cedric Driesen}%
\affiliation{Hasselt University, B-3590 Diepenbeek, Belgium.}%
\author{Bart Cleuren}
\affiliation{Hasselt University, B-3590 Diepenbeek, Belgium.}%
\author{Christian Van den Broeck}
\affiliation{Hasselt University, B-3590 Diepenbeek, Belgium.}%

\date{\today}% It is always \today, today,

\begin{abstract}
We study the efficiency of a single particle  Szilard and Carnot engine. Within a first order correction to the quasi-static limit, the work distribution is found to be Gaussian and the correction factor to average work and efficiency only depends on the piston speed. The stochastic efficiency is studied for both models and the recent findings on efficiency fluctuations are confirmed numerically. Special features are revealed in the zero temperature limit.
\end{abstract}

\pacs{05.70.Ln, 07.20.Pe, 02.50.Ey}% PACS, the Physics and Astronomy
\maketitle

%\tableofcontents
\section{Introduction}
Single particle engines have long been regarded as thought experiments, designed to probe the very foundations of thermodynamics and statistical mechanics. A celebrated example is the Szilard engine  \cite{szilard_uber_1929,parrondo2015thermodynamics,maruyama2009colloquium}, conceived to investigate the notion of information in thermodynamics. By now, with the spectacular progress in bio- and nanotechnology, these machines are part of an experimentally accessible reality \cite{berut_experimental_2012,toyabe_experimental_2010,gieseler2014dynamic,wang2002experimental,blickle2006thermodynamics,PhysRevLett.114.120601,martinez_brownian_2014}. Alongside these developments, there has been significant progress on the theoretical side, starting with the discovery of the fluctuation theorem and the Jarzynski and Crooks relations \cite{evans_probability_1993,jarzynski_nonequilibrium_1997,crooks_entropy_1999}, and developing into encompassing theories such as stochastic thermodynamics \cite{seifert_stochastic_2012,van_den_broeck_ensemble_2014}. An intriguing recent discovery concerns the properties of the stochastic efficiency \cite{verley_unlikely_2014,verley_universal_2014,proesmans_stochastic_2015,Proesmansfcs,gingrich_efficiency_2014,polettini_efficiency_2015,rana_single-particle_2014,PhysRevB.91.115417,jiang2015efficiency,rana2015anomalous,martinez_brownian_2014}. This new concept derives from the observation that the performance, and hence also the efficiency of small machines is subject to strong fluctuations. Based upon the fluctuation theorem, it was demonstrated that for asymptotically long times,  the reversible efficiency is least probable for a machine operating under a time-symmetric schedule, while under a time-asymmetric driving the probability distributions of efficiency of the time-forward and the time-reversed protocol intersect at the reversible efficiency.

In the present paper we investigate a generic set-up for a stochastic engine, namely a single particle moving between a thermal wall and a piston \cite{bena2005jarzynski,lua2005practical,baule2006validation,izumida2008molecular,hooyberghs2013efficiency,hoppenau2013carnot,itami2014nonequilibrium,sano2014stochastic,cerino2015kinetic}, and apply the obtained results to  the Szilard engine and a simplified Carnot heat engine. For both engines we investigate the macroscopic efficiency and the efficiency fluctuations. 
The paper is organised as follows. Section \ref{genSetup} introduces the Szilard engine and the simplified Carnot engine. While the principle behind each engine is different, i.e. information to work conversion and heat to work conversion, respectively, the constitutive process, namely a particle moving between a stationary thermal wall and a perfectly reflecting moving wall, is the same. In section \ref{quasi} we analyse the quasi-static operation including a first order correction. In this limit, the work probability distribution for the constitutive process is Gaussian, analogous to other isothermal processes \cite{speck2004distribution,speck2011work,nickelsen2011asymptotics,hoppenau2013work}, and the average work is the free energy difference times a factor that only depends on the reduced velocity defined as the ratio of the velocity of the piston and the thermal velocity associated with the thermal wall. The implications for both engines are discussed. In section \ref{efflux}, we present a detailed numerical analysis  of efficiency fluctuations, including finite time results and confirming the predicted universal properties in the asymptotic time limit. A slightly modified heat engine is introduced for which analytical results are possible. Conclusions and outlook are given in section \ref{sum}.
%%%%%%%%%%%%%%%%%%%%%%%%%%%%%%%%%%%%%%%%%%%%%%%%%%%%%%%%%%%%%%%%%%%%%%%%%%
\section{General set-up}\label{genSetup}
\begin{figure}\begin{centering}
\includegraphics[width=\columnwidth]{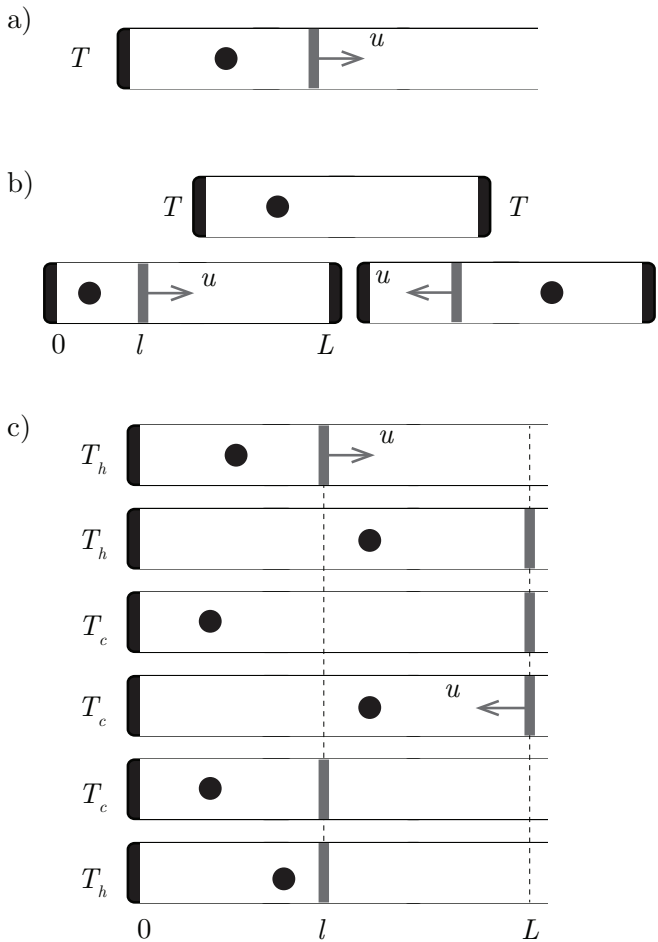}
\caption{Schematic representation of: a) The constitutive process: a single particle bounces back and forth between a thermal wall on the left and a moving piston on the right. b) The Szilard engine: a piston is inserted a distance $l$ from the left wall, and subsequently moves away from the particle. c) A simplified Carnot heat engine: alternating cycles of expansion and compression, while in contact with a hot or cold left wall at temperatures $T_h$ and $T_c$ respectively; note the absence of an adiabatic expansion and compression phase.\label{fig1}}
\end{centering}\end{figure}

The various heat and work exchange processes are schematically represented in figure \ref{fig1}. We consider, without loss of generality, a one-dimensional setting.  The constitutive process is shown in figure \ref{fig1}a and corresponds to a single particle with mass $m$ moving freely between a stationary thermal wall and an infinitely heavy moving piston. 
Two types of thermodynamic events occur: either the particle collides with the thermalising wall, or it collides with the piston. In case of a collision with the thermalising wall, the particle gets a new velocity drawn from the Rayleigh distribution \cite{alec}:
\begin{equation}\phi(v)=v e^{-\frac{ v^{2}}{2}},\label{v0dis}\end{equation}
with $v$ the reduced, dimensionless velocity of the particle, measured in terms of the thermal velocity $\sqrt{k_BT/m}$ with $k_B$ the Boltzmann constant and $T$ the temperature of the wall. The change in kinetic energy then corresponds to an exchange of heat with the thermal reservoir. Collisions with the piston are assumed purely elastic. Hence a particle with incoming velocity $v$, hitting the piston moving at (dimensionless) velocity $u$, will recoil with velocity $2u-v$. The change in kinetic energy of the particle implies an amount of work $2k_BTu(v-u)$ done on the piston.
%%%%%%%%%%%%%%%%%%%%%%%%%%%%%%%%%%%%%
\subsection{The Szilard engine}
For the Szilard engine, shown in figure \ref{fig1}b, the various steps per cycle are as follows. At the start, a particle bounces back and forth between two thermal walls at temperature $T$ and separated by a distance $L$. The  probability distribution for its position $x_0$ and  (dimensionless) velocity $v_0$ is given by thermal equilibrium:
\begin{equation}
p_{L}(x_0,v_0)=\frac{1}{L}\frac{1}{\sqrt{2\pi}}\exp\left(-\frac{v_0^2}{2}\right).\label{eq}
\end{equation}
A piston is then inserted a distance $l$ from the left wall, trapping the particle either on its left or right hand side. A measurement is performed to find out in which compartment the particle resides. If the particle is in the left compartment, the thermodynamic cost of the measurement is: $k_B\Delta i=-k_B\ln x$ \cite{cover2012elements}, where $x=l/L$ is the compression ratio. The equilibrium distribution is now restricted to the left compartment, i.e.~given by (\ref{eq}) but with $L$ replaced by $l$. The piston next moves to the right with velocity $u$ until the full length $L$ is reproduced, and an amount of work $w$ is delivered. For  finite piston speed, the system will deviate from  equilibrium  and the average of this work is bounded by the decrease $\Delta F$ in free energy,  $\langle w\rangle \leq \Delta F$ with:
\begin{equation}\label{F}
\Delta F=-k_BT\ln x
\end{equation}
When the particle is found in the right compartment, analogous results hold for the measurement cost $\Delta i$, the decrease in free energy $\Delta F$ and the work $w$, by replacing $x$ with $1-x$. The efficiency of the above information to work conversion is defined as:
\begin{equation}\eta=\frac{w}{k_B T\Delta i}.\end{equation}
We will investigate the properties of this stochastic quantity in more detail below. Upon repeating the Szilard cycle many times, $\eta$ will converge to a macroscopic efficiency $\bar{ \eta}$, which is, as we show below, bounded by the reversible efficiency $1$:
  \begin{equation}
 \bar{ \eta}=\frac{\left\langle w\right\rangle}{k_B T \left\langle \Delta i\right\rangle}\leq 1,
 \end{equation}
see also figure \ref{figME}a.
Referring to the discussion of stochastic efficiency,  we also require the time-reversed process of the Szilard engine, which is defined as follows. Initially, the particle is in thermal equilibrium and moving between the two thermalising walls. The piston is then brought in from one of the sides and pushed inward up to position $l$. The piston either starts at position $0$, with probability $1-x$, or at position $L$ probability $x$. In both cases, work is converted into information.
%%%%%%%%%%%%%%%%%%%%%%%%%%%%%%%%%%%%%
\subsection{The simplified Carnot heat engine}
The various stages of the single particle Carnot heat engine are represented in figure \ref{fig1}c.  When discussing the heat engine, the reduced velocities are calculated with respect to the thermal velocity at temperature $T_h$. The cycle starts with the piston moving outward (stage 1) at constant velocity $u$ until it reaches its final position $L$. The particle is allowed to relax (stage 2), i.e.~the equilibrium distribution Eq.~({\ref{eq}) is restored. After relaxation, the temperature of the heat bath is instantaneously switched to the lower value $T_c<T_h$, and the particle is again allowed to relax to the new equilibrium state (stage 3). Next (stage 4) the piston moves inwards with velocity $u$ until it has returned to its original position $l$, after which the particle again relaxes (stage 5).  Finally, the temperature is instantaneously switched back to the higher value $T_h$, followed by a final relaxation (stage 6). The inclusion of the intermediate relaxation stages destroys any correlations in position and velocity that would otherwise appear between the successive stages, the relaxation stages are purely dissipative, therefore decreasing the efficiency, and have an infinitely long duration, which leads to zero power output. The whole cycle is then repeated.  In this way heat $q$ from the hot bath is converted into work $w$ with stochastic efficiency:
\begin{equation}\eta=\frac{w}{q}.\label{etaHE}\end{equation}
The macroscopic efficiency is bounded by:
 \begin{equation}
\bar{\eta}=\frac{\left\langle w\right\rangle}{\left\langle q\right\rangle}\leq \eta_C=1-\frac{T_c}{T_h}.
\end{equation} This bound is saturated only in the quasi-static regime $u \rightarrow 0$, with the additional limit $x= l/L\rightarrow 0$. For large $u$ on the other hand, the set-up will no longer operate as a heat engine, since the average amount of delivered work becomes negative.
The time-reversed process corresponds to an expansion of the piston, while being in contact with the cold reservoir, followed by a compression while in contact with the hot reservoir. Under this protocol, the system operates as a refrigerator.

\begin{figure}\begin{centering}
\includegraphics[width=\columnwidth]{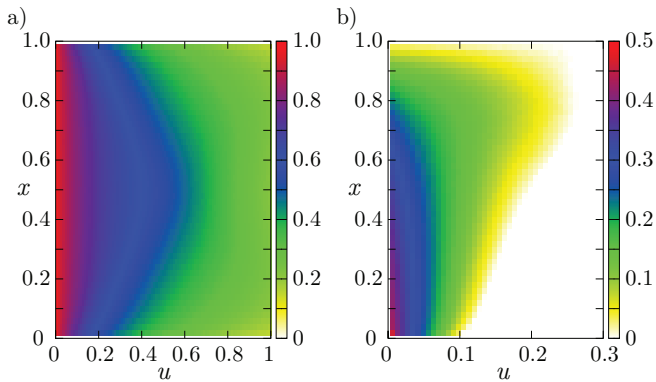}
\caption{(Color
 online) Macroscopic efficiency of a) the Szilard engine and b) the heat engine (with $T_h=2T_c$) in function of $u$ and $x$.\label{figME}}
\end{centering}\end{figure}
%%%%%%%%%%%%%%%%%%%%%%%%%%%%%%%%%%%%%%%%%%%%%%%%%%%%%%%%%%%%%%%%%%%%%%%%%%
\section{Quasi-static limit}\label{quasi}
We first focus on the quasi-static limit of the  constitutive process, and consider the work delivered by the particle, initially at equilibrium and in presence of the thermal wall  at temperature $T$, on a  piston, moving with dimensionless  speed $u$ from  the initial  $x=l/L$ to  final (reduced) location $1$.  The quasi-static regime corresponds to the limit in which the velocity goes to zero, $u \rightarrow 0$. At lowest order, one obtains the familiar result that the work converges to the decrease $\Delta F$ in free energy. An interesting feature appears when considering   the first correction, which turns out to be of order $u\ln u$. In appendix A, it is shown that the probability  distribution $p_{x,u}(w)$ for the work $w$, upon moving the piston from  $x$ to  $1$ at constant velocity $u$, is, at this order, given by a Gaussian distribution, cf. figure \ref{fig5}:
\begin{equation}p_{x,u}(w)=\frac{1}{\sqrt{2\pi} \sigma_{x,u}}e^{-\frac{(w-\left\langle w\right\rangle_{x,u})^2}{2\sigma_{x,u}^2}}+O(u).\label{Compeq}\end{equation}
The convergence to the Gaussian entails that all cumulants of higher than second order decrease proportional to $u$ in the limit $u\rightarrow 0$:
\begin{equation}\ln\left\langle e^{\lambda w}\right\rangle=\left\langle w\right\rangle_{x,u}\lambda+\sigma_{x,u}^2\frac{\lambda^2}{2}+O(u).\end{equation}
Remarkably, the reduction in average work due to the finite speed of the piston only depends on the reduced speed:
\begin{equation}
\left\langle w\right\rangle_{x,u}=\Delta F\left(1+\sqrt{\frac{2}{\pi}}u\ln u\right).\label{QSMeff}
\end{equation}
Similar for the dispersion:
\begin{equation}\sigma_{x,u}^2=-2k_B T \Delta F\sqrt{\frac{2}{\pi}}u\ln u.\label{QSSig}\end{equation}
We stress that the $u\ln u$ correction stems from the fact that the average time for a particle, initialised in the equilibrium state, to reach the thermal wall diverges logarithmically as $u\rightarrow 0$. We note that this result differs from the linear dependence found in e.g.~low dissipative systems \cite{esposito2010efficiency}.  Furthermore, the convergence to a Gaussian distribution is rather slow, as the non-Gaussian corrections of order $u$ only become negligible compared to $u\ln u$ for very small values of $u$.

The corresponding result for the time-reversed process, i.e.~compression of the piston from position $1$ to position $x$ with dimensionless velocity $u$, is obtained by replacing  $u\ln u$ with $-u\ln u$ and $\Delta F$  by $-\Delta F$. Hence the corresponding probability distribution, denoted by $\tilde{p}_{x,u}(w)$, is also Gaussian with first two central moments 
\begin{eqnarray}\label{mi}
\tilde{\left\langle w\right\rangle}_{x,u}&=&\left\langle w\right\rangle_{x,u}-2\Delta F\\
\tilde{\sigma}_{x,u}&=&\sigma_{x,u}. 
\end{eqnarray}
We are thus in a position to perform an independent check of the above finding, namely by verifying the   Crooks relation \cite{crooks_entropy_1999}. Since
\begin{eqnarray}\frac{p_{x,u}(w)}{\tilde{p}_{x,u}(-w)}&=&\exp\left[-\frac{(w-\left\langle w\right\rangle_{x,u})^2-(-w-\tilde{\left\langle w\right\rangle}_{x,u})^2}{2\sigma_{x,u}^2}\right]\nonumber\\&=&\exp\left[\frac{\left\langle w\right\rangle_{x,u}+\tilde{\left\langle w\right\rangle}_{x,u}}{\sigma_{x,u}^2}(w-\Delta F)\right],\end{eqnarray}
we find that the Crooks relation implies:
\begin{equation}\frac{\left\langle w\right\rangle_{x,u}+\tilde{\left\langle w\right\rangle}_{x,u}}{\sigma_{x,u}^2}=-\frac{1}{k_B T},\end{equation}
which is indeed in agreement with Eqs. (\ref{QSMeff}), (\ref{QSSig}) and ({\ref{mi}).

%%%%%%%%%%%
\subsection{The Szilard engine}
For the Szilard engine, the distribution of information gain and work during one iteration can be written as:
\begin{multline}
P_1(\Delta i,w)=xp_{x,u}(w)\delta(\Delta i+\ln x)\\
+(1-x)p_{1-x,u}(w)\delta (\Delta i+\ln(1-x)).
\end{multline}
 The probability distribution of the stochastic efficiency of one iteration can be written as:
\begin{multline}
P_1(\eta)=xp_{x,u}(-k_BT\eta\ln x)\\+(1-x)p_{1-x,u}(-k_BT\eta\ln(1-x)).\label{PetaSzil}
\end{multline}
Combining with Eq.~(\ref{Compeq}), we conclude that the probability distribution of the efficiency of the Szilard engine in the quasi-static limit can be written as a sum of two Gaussian distributions.
Furthermore, the macroscopic efficiency is found to be, cf. Eq.~(\ref{QSMeff}):
\begin{eqnarray}\bar{\eta}&=&-\frac{x\left\langle w\right\rangle_{x,u}+(1-x)\left\langle w\right\rangle_{1-x,u}}{k_B T\left(\ln x+\ln(1-x)\right)}\nonumber\\&=&1+\sqrt{\frac{2}{\pi}}u\ln u +O(u).\end{eqnarray}
Note that the first order correction to the macroscopic efficiency is independent of the initial position of the piston, which leads to universal behaviour in the efficiency near the quasi-static limit, as can be seen in figure \ref{fig6}a. 
%%%%%%%%%%%
\subsection{The simplified Carnot heat engine}
The above calculations can also be repeated for the heat engine. The joint probability distribution of produced work and absorbed heat from the hot reservoir in one cycle can be written as:
\begin{multline}
P_1(w,q)=\\
\int dw_0\,p^{h}_{x,u}(w_0)\,\tilde{p}^{c}_{x,u\sqrt{\frac{T_h}{T_c}}}(w-w_0)\,f(q-w_0).
\end{multline}
The index $h$ or $c$ appearing in the probability distribution of the constitutive process refers to the temperature of the thermal wall ($T_h$ and $T_c$, respectively) during that stage of the cycle. The extra factor $\sqrt{T_h/T_c}$ in the subscript of $\tilde{p}^{c}$ accounts for the fact that the velocity is always relative to the thermal velocity of the wall, and the factor $f(q-w_0)$ accounts for the change in kinetic energy upon switching the temperature of the heat bath at the end of stages 2 and 5. The corresponding probability distribution $f(q)$ is given by:
\begin{multline}f(q)=\\
\iint dv_h\, dv_c\,\frac{\sqrt{T_h}}{2\pi T_c}e^{-\frac{1}{2}\left(v_h^2+\frac{T_h}{T_c}v_c^2\right)}\delta\left(q-k_B\frac{T_hv_h^2-T_cv_c^2}{2}\right)\\
=\frac{e^{\left(\frac{1}{k_BT_c}-\frac{1}{k_BT_h}\right)\frac{q}{2}}}{\pi k_B\sqrt{T_hT_c}}K_0\left(\left(\frac{1}{k_BT_c}+\frac{1}{k_BT_h}\right)\frac{\left|q\right|}{2}\right),
\end{multline}
with $K_0$ the modified Bessel function of the second kind \cite{olver2010nist}.
Although there appears to be no simple expression for the probability distribution of the stochastic efficiency $\eta=w/q$, we can calculate analytically the macroscopic efficiency in the quasi-static limit, cf. appendix A:
\begin{multline}\bar{\eta}=\frac{2(T_h-T_c)\ln x}{2T_h\ln x+T_c-T_h}\\+\frac{2\sqrt{2}\ln x}{\sqrt{\pi}}\frac{(2\sqrt{T_hT_c}\ln x+T_c-T_h)(T_h+\sqrt{T_hT_c})}{(2 T_h\ln x+T_c-T_h)^2}u\ln u\\ +O(u)\label{etahb}.\end{multline}
This is confirmed by figure \ref{fig6}b. As mentioned earlier, Carnot efficiency can be reached in the limit $x\rightarrow 0$:
\begin{equation}\bar{\eta}=1-\frac{T_c}{T_h}+\sqrt{\frac{2}{\pi}}\frac{T_c}{T_h}\left(1+\sqrt{\frac{T_h}{T_c}}\right)u\ln u+O(u),\end{equation}
see also figure \ref{figME}b.

\begin{figure}\begin{centering}
\includegraphics[width=\columnwidth]{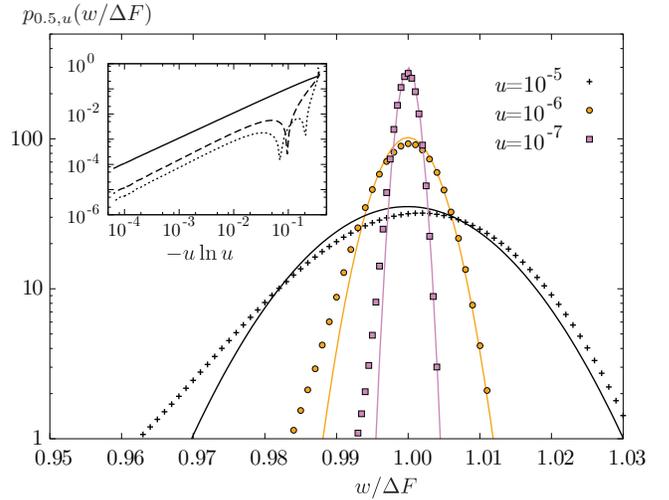}
\caption{(Color
 online) Gaussian approximation (solid lines) of the work distribution for an expansion from $l=0.5$ to $L=1$ near the quasi-static limit, together with numerical results, for $u=10^{-5}$ (black crosses), $10^{-6}$ (orange dots) and $10^{-7}$ (purple squares). Inset: second (solid line), third (dashed line) and fourth (dotted line) cumulant of the scaled work distribution.\label{fig5}}
\end{centering}\end{figure}

\begin{figure}
\includegraphics[width=\columnwidth]{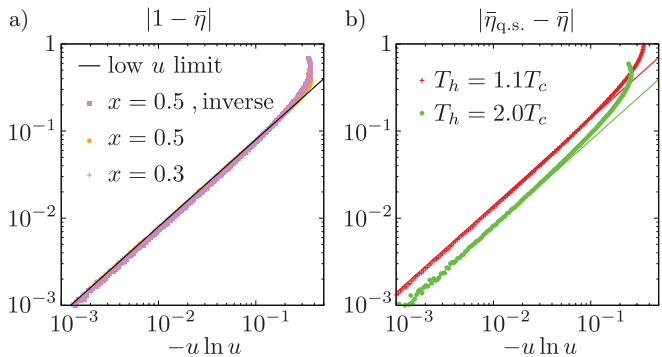}
\caption{(Color online) Macroscopic efficiency of a) the Szilard engine near the quasi-static limit, with $x=0.7$ (light blue), $x=0.5$ (orange) and the time-inverse process of $x=0.5$ (purple), and b) the heat engine near the quasi-static limit with $x=0.5$ for $T_h=1.1T_c$ (red) and $T_h=2T_c$ (green).\label{fig6}}
\end{figure}
%%%%%%%%%%%%%%%%%%%%%%%%%%%%%%%%%%%%%%%%%%%%%%%%%%%%%%%%%%%%%%%%%%%%%%%%%%
\section{Efficiency fluctuations}\label{efflux}
\begin{figure}
\includegraphics[width=\columnwidth]{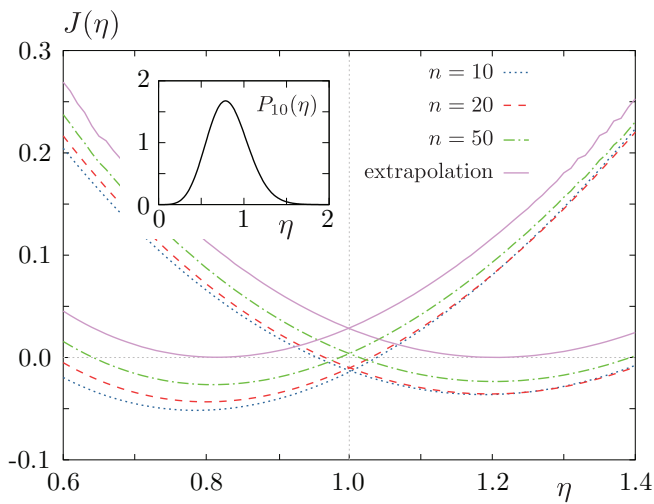}
\caption{(Color online) $-(1/n) \ln P_n(\eta)$ for $10$ (blue dotted line), $20$ (red dashed line) and $50$ (dashed dotted green line) cycles of the Szilard engine and its time-inverse, with $u=0.1$ and $x=0.7$. The purple full curve is the extrapolation to the LDF. The macroscopic efficiency is $\bar{\eta}=0.80$. The inset shows $P_{10}(\eta)$. \label{fig2}}
\end{figure}
We now turn to the efficiency fluctuations of the engines. The focus here is mainly on the large deviation function (LDF) $J(\eta)$ of the probability distribution $P_n(\eta)$ of the stochastic efficiency \cite{touchette_large_2009}:
\begin{equation}J(\eta)=-\lim_{n\rightarrow \infty}\frac{1}{n}\ln P_n(\eta),\end{equation}
with $n$ the number of cycles of the engine. For systems under time-asymmetric driving, the LDF  of the time-forward process $J(\eta)$ and the LDF of the time-reversed process $\tilde{J}(\eta)$ intersect at reversible efficiency $\eta_r$ \cite{verley_universal_2014,gingrich_efficiency_2014}:
\begin{equation}J(\eta_r)=\tilde{J}(\eta_r).\label{EtaLDFRev}\end{equation}
In our numerical simulations, it is of course impossible to take the limit of an infinite number of cycles $n \rightarrow \infty$. However the convergence to this asymptotic limit is in itself an interesting feature. We include the estimate of the large deviation function using an extrapolation scheme, which is presented in detail in \cite{proesmans_stochastic_2015,Proesmansfcs}, and which leads to accurate results. 

The approach to the LDF for the Szilard engine are shown in figure \ref{fig2}, including the  result of the aforementioned extrapolation scheme. The obtained estimate of the efficiency LDF reproduces the expected results: $J(\eta)\geq 0$ and reaches zero at macroscopic efficiency; the LDF of forward and the reversed process intersect at reversible efficiency $\eta_r$.  Note however that the present model differs from the theory discussed in \cite{verley_universal_2014,proesmans_stochastic_2015,polettini_efficiency_2015}, because  the information gain/loss in the engine cannot become zero, even at a stochastic level. Consequently, the LDF has no finite plateau at infinity and no power-law behaviour is found in the corresponding probability distribution.  

The approach to the LDF of the time-forward and the time-reversed heat engine are plotted in figure \ref{fig3}. The LDFs intersect at Carnot efficiency, as expected. The other predictions of the general analysis \cite{verley_universal_2014} are reproduced: the maxima of both LDFs have the same height $J(\eta)$ converges to the same finite plateau value for $\eta \rightarrow \pm \infty$,  and  similarly so for the time-reversed process. Concerning the finite time regime, we note that the crossing of the probability distributions is actually always quite close to the reversible efficiency and  seems to converge to Carnot efficiency as a power law in function of the number of cycles, cf.  inset of figure \ref{fig3}.

In order to derive analytical results for the efficiency fluctuations, we introduce a slightly modified version of the heat engine, and take the limit $T_c \rightarrow 0$. When the temperature of the thermal wall is zero, the particle sticks to it upon collision. The modification is that the hot reservoir at temperature $T_h$ is present only during an infinitesimal short time at the start of the cycle. In this way, the particle is launched with a velocity $v$ drawn from the Rayleigh distribution (Eq.~(\ref{v0dis})) at temperature $T_h$. Immediately following the launch, the temperature of the thermal wall is dropped to zero. Keeping the relaxation step after the expansion of the piston, the particle eventually returns to the thermal wall and the piston can be returned to its initial position without delivering any work. It is clear that the relaxation step at the end of the cycle becomes irrelevant, as the particle is immobilised, and can be discarded. With this time-symmetric protocol, the particle will collide exactly once with the piston during each cycle. Furthermore, setting $x = 0.5$, the particle delivers work only when $v > 2u$. The efficiency distribution during one cycle is given by:
\begin{widetext}\begin{eqnarray}P_{1}(\eta)&=&\int dv\, \phi(v)\left(\theta(2u-v)\delta(\eta)+\theta(v-2u)\delta\left(\eta-\frac{4u(v-u)}{v^2}\right)\right)\nonumber\\&=&\left(1-e^{-2 u^2}\right)\delta(\eta)+\frac{2 u^2}{\eta^3}\frac{f(\eta)^4e^{-2\left(\frac{f(\eta)u}{\eta}\right)^2}}{f(\eta)-\eta}\theta(1-\eta^2),\label{PetaT0}\end{eqnarray}
\end{widetext}
with $f(\eta)=1+\sqrt{1-\eta}$ and $\theta(x)$ the Heaviside function. The LDF can be found by numerically contracting the cumulant generating function, as is done in the appendix.
The results are shown in figure \ref{fig4}. It is clear that, even at a stochastic level, only efficiencies between $0$ and $1(=\eta_C)$ can be reached. Furthermore, the LDF diverges at Carnot efficiency and the results from the extrapolation are in good agreement with the analytical formula derived in appendix B.

\begin{figure}\begin{centering}
\includegraphics[width=\columnwidth]{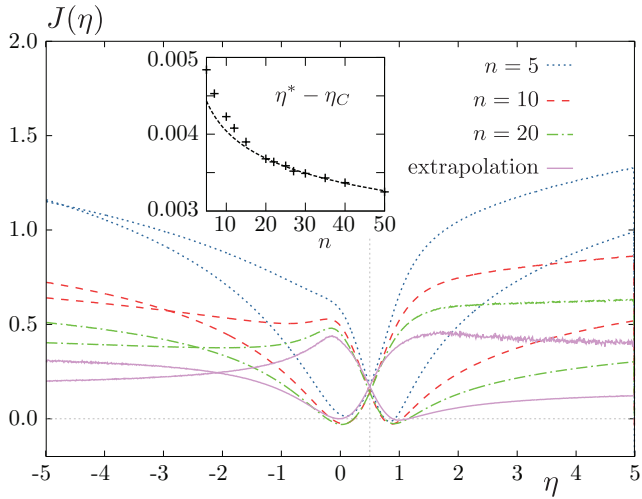}
\caption{(Color
 online) $-(1/n) \ln P_n(\eta)$ for $5$ (blue dotted line), $10$ (red dashed line) and $20$ (green dashed, dotted line) cycles of the heat engine and its time-inverse, with $T_h=2T_c$ (i.e.~$\eta_C=1/2$), $u=0.3$ and $x=0.5$. The purple full curve is the extrapolation to the LDF. The macroscopic efficiency is given by $\bar{\eta}=-0.02$ Inset: convergence of the intersections efficiency $\eta^{*}$ of forward and time-reverse curves to $\eta_C$ as the number $n$ of cycles increases. The dashed line is a power law fit of the form $\alpha/n^{\beta}$, with $\alpha=5.49\cdot 10^{-3}$ and $\beta=0.13$\label{fig3}}
\end{centering}\end{figure}

\begin{figure}\begin{centering}
\includegraphics[width=\columnwidth]{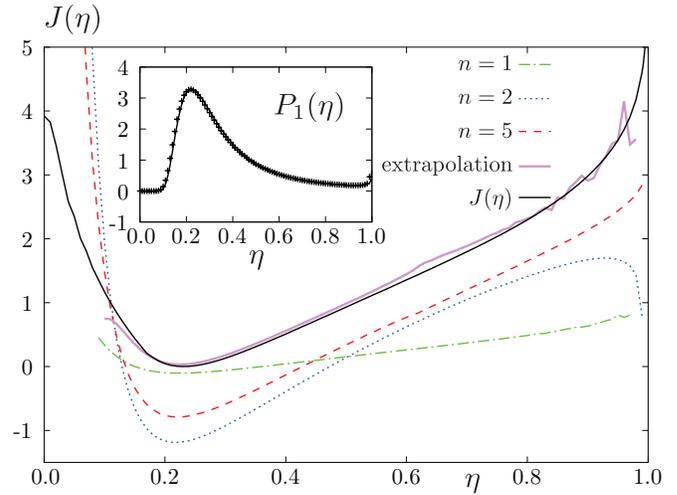}
\caption{(Color online) $-(1/n) \ln P_n(\eta)$ for $1$ (blue dotted line), $2$ (red dashed line) and $5$ (green dashed, dotted line) cycles of the zero-temperature heat engine, with $u=0.1$. The purple full curve is the extrapolation to the LDF and the black curve is the analytical LDF. The macroscopic efficiency is given by $\bar{\eta}=0.23$. Inset: $P_1(\eta)$.\label{fig4}}
\end{centering}\end{figure}
\section{Discussion}\label{sum}
We have analysed the performance of two typical single particle engines, the Szilard engine and a simplified heat engine. Their analysis is based on the dynamics of a single particle moving back and forth between a thermalising wall and a moving piston. For both engines, we have shown that the distribution of delivered work during one cycle becomes Gaussian in the quasi-static limit. More precisely, the corrections of the first two cumulants in the quasi-static limit scale as $u\ln u$, whereas the higher order cumulants are proportional to $u$. 
For the heat engine,  both work $w$ and heat $q$ can take on any real value, and the fluctuations of  efficiency $\eta$  reproduce the familiar universal features. For the Szilard engine however, the information gain per cycle can only take on two (positive) values $-\ln x$ or $-\ln(1-x)$. As a result, the asymptotes of the LDF's for the efficiency diverge, since extreme (positive or negative) efficiencies can only arise from a corresponding extreme value of the delivered work. On the other hand, the universal crossing of forward and time-reverse LDF's at reversible efficiency $1$ is reproduced. 
Our results can easily be extended to other systems such as the full Carnot cycle \cite{hoppenau2013carnot} and the Andersen cycle \cite{cerino2015kinetic}. A question for further research  is to clarify how the sub-linear $u\ln u$ behavior fits into the linear thermodynamics description of piston engines \cite{izumida2008molecular}. 

\begin{acknowledgments}
The computational resources and services used in this work were provided by the VSC (Flemish Supercomputer Center), funded by the Hercules Foundation and the Flemish Government – department EWI
\end{acknowledgments}

\appendix

\section{Quasi-static limit}
In \cite{hoppenau2013carnot}, it is shown that the probability distribution $p_{x,u}(w,\tau)$ of the produced work, with $\tau=\ln\left(1+ut/L_0\,\sqrt{k_BT/m}\right)$, $L_0$ being the initial position of the piston, $t$ the time, $T$ the temperature, and $m$ the mass of the particle, is given by:
\begin{equation}\hat{p}_{x,u}(\omega,\lambda)=\frac{\hat{p}_{i;x,u}(\omega,\lambda)\hat{p}_{f;x,u}(\omega,\lambda)}{1-\hat{\psi}_{u}(\omega,\lambda)}+O(u),\end{equation}
with $\hat{p}_{x,u}(\omega,\lambda)$ the probability distribution of the delivered amount of work, with $w$ Fourier transformed to $\omega$ and $\tau$ Laplace transformed to $\lambda$. Furthermore,
\begin{equation}\hat{p}_{i;x,u}(\omega ;\lambda)=1-\lambda\left\langle\tau_0\right\rangle_{x,u}+i\omega\left\langle w_0\right\rangle_{x,u}+O(u),\end{equation}
\begin{equation}\hat{\psi}_{u}(\omega,\lambda)=e^{2ik_BT\omega u^2}g_{2u}(\omega,\lambda),\end{equation}
\begin{multline}\hat{p}_{f;x,u}(\omega;\lambda)=\\\frac{1}{\lambda}\left(1-g_{u}(0,\lambda)+e^{2ik_BT\omega u^2}g_{u}(2\omega,\lambda)-\hat{\psi}_u(\omega;\lambda)\right),\end{multline}

\begin{eqnarray}g_{u}(\omega,\lambda)&=&\int^{\infty}_{\max(u,0)}dv\,\phi(v)\left(1-\frac{u}{v}\right)^{\lambda}e^{-i\omega k_BTuv}\nonumber\\&=&1-u\sqrt{\frac{\pi}{2}}\lambda-iu\sqrt{\frac{\pi}{2}}\omega k_BT\nonumber\\&&-\frac{\lambda(\lambda-1)}{2} 
u^2\ln u+O(u^2),\end{eqnarray}
with $\left\langle\tau_0\right\rangle_{x,u}$ the average modified time before the first collision with the thermal wall, $\left\langle w_0\right\rangle_{x,u}$ the average delivered amount of work after the first collision with the thermal wall, and $\phi(v)$ the Rayleigh distribution. Filling in the last equation in the upper two, gives:
\begin{multline}\hat{\psi}_u(\omega;\lambda)=1-\sqrt{2\pi}u \left(\lambda+ik_BT \omega\right)\\-2\lambda(\lambda-1)u^2\ln u+O(u^2),\end{multline}
and
\begin{equation}\hat{p}_{f;x,u}(\omega;\lambda)=u\sqrt{2\pi}+2(\lambda-1)u^2\ln u+O(u^2).\end{equation}
If the system is initially in its steady state, we have:
\begin{equation}\left\langle\tau_0\right\rangle_{x,u}=-\sqrt{\frac{2}{\pi}}u\ln u+O(u),\label{avtime}\end{equation}
\begin{equation}\left\langle w_0\right\rangle_{x,u}=O(u),\end{equation}
so that:
\begin{equation}\hat{p}_{i;x,u}(\omega;\lambda)=1+\lambda\sqrt{\frac{2}{\pi}}u\ln u+O(u).\end{equation}
Combining these results, leads to:
\begin{multline}
\hat{p}_{x,u}(\omega;\lambda)=\frac{1}{\lambda+i\omega k_BT}\\
+\sqrt{\frac{2}{\pi}}\frac{2i\omega\lambda k_BT+\lambda^2-i\omega k_BT}{(\lambda+i\omega k_BT)^2}u\ln u+O(u).\end{multline}
The probability distribution of the produced amount of work $w$ and the information consumption $\Delta i$ can than be directly calculated by inverse Laplace and Fourier transforming. Furthermore, Eq.~(\ref{avtime}) shows that the average time before the first collision diverges logarithmically when $u\rightarrow 0$.

\begin{figure}\begin{centering}
\includegraphics[width=\columnwidth]{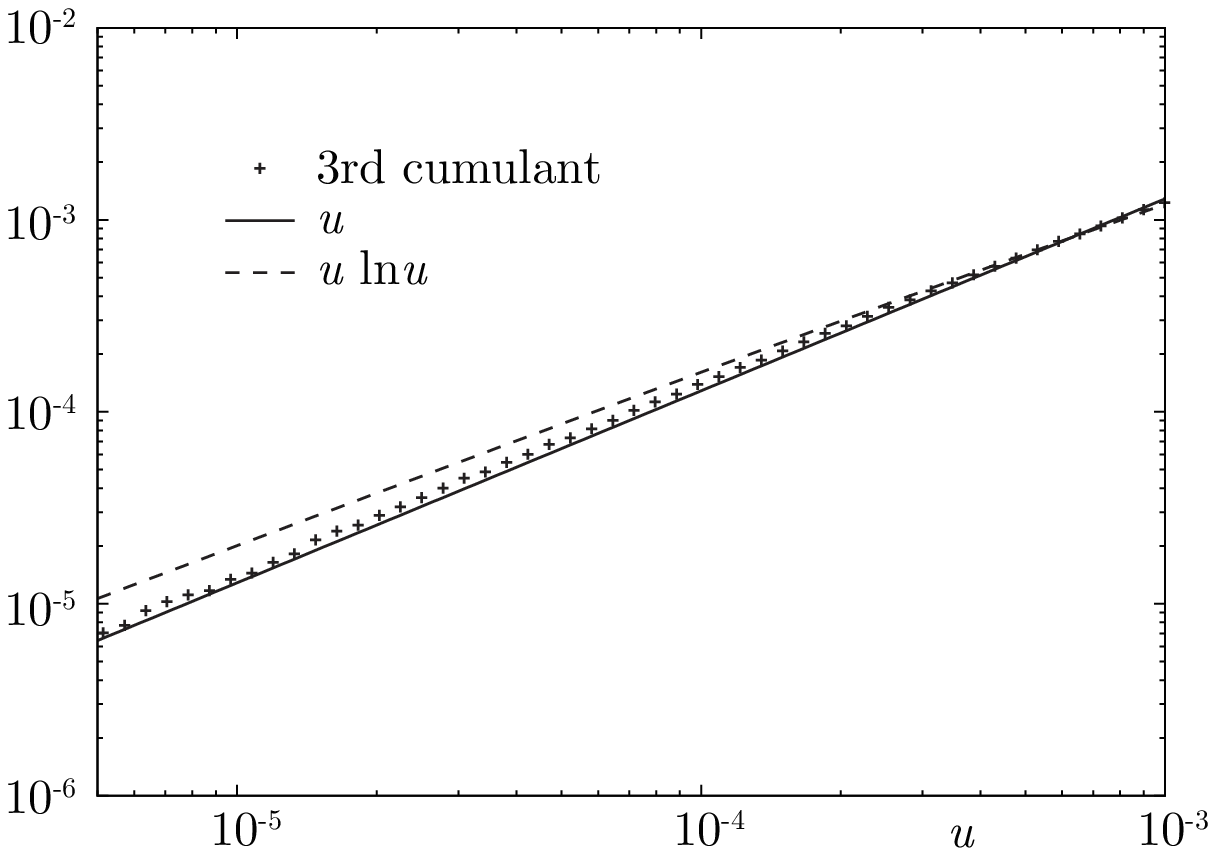}
\caption{$3$rd cumulant of the work distribution in function of $u$ for $x=0.5$, with $u$ and $u\ln u$ fit. \label{fig3Cum}}
\end{centering}\end{figure}
\begin{figure}\begin{centering}
\includegraphics[width=\columnwidth]{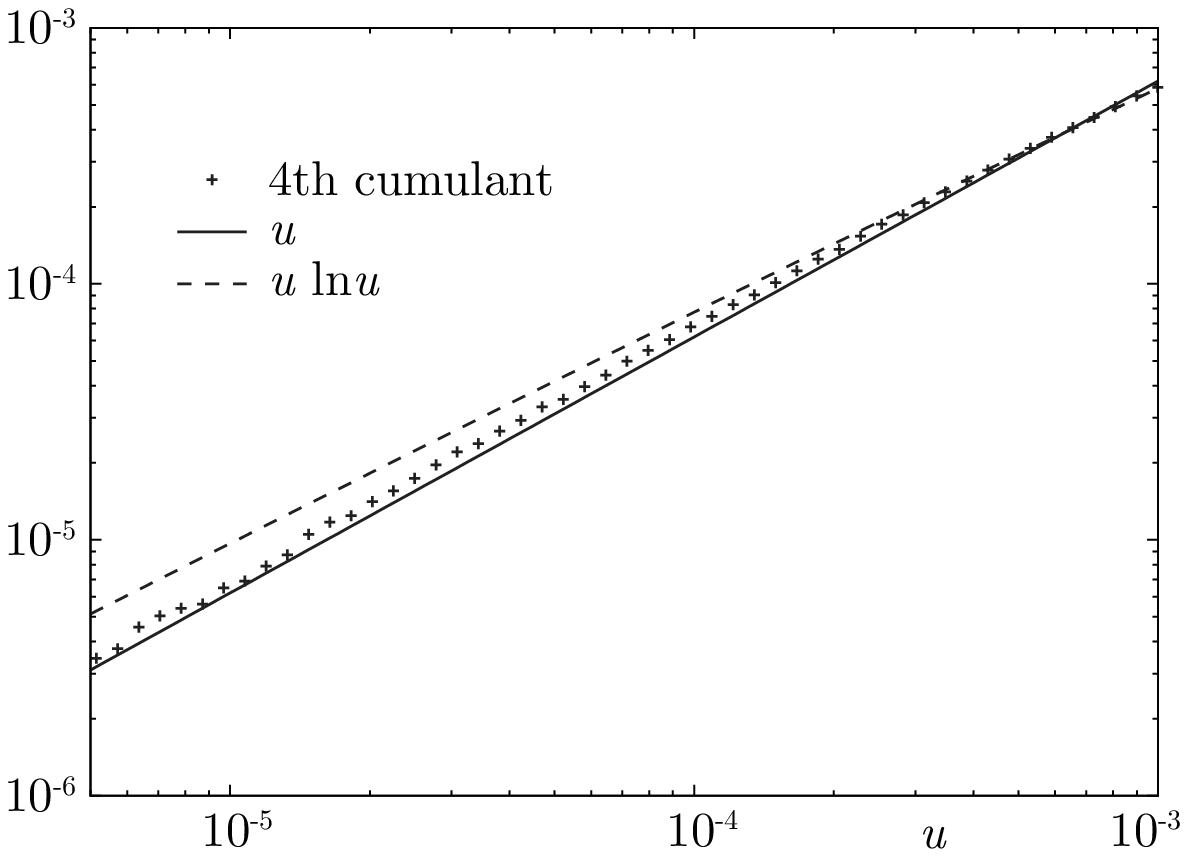}
\caption{$4$th cumulant of the work distribution in function of $u$ for $x=0.5$, with $u$ and $u\ln u$ fit. \label{fig4Cum}}
\end{centering}\end{figure}
To numerically check that the higher order cumulants of the work distribution go as $u$ instead of $u\ln u$, we plot the third and fourth cumulant of the distribution of $ w/\Delta F$ in function of $u$ and fit to both $u$ and $u\ln u$. These plots clearly show that those cumulants are proportional to $u$, as was discussed above.

It is also possible to calculate the efficiency of the heat engine. The delivered amount of work can be directly found from the above calculation:
\begin{multline}\left\langle w\right\rangle_{x,u}=-k_B\ln x \left(T_h-T_c\right)\\-\sqrt{\frac{2}{\pi}}k_B\ln x \left(T_h+\sqrt{T_hT_c}\right)u\ln u+O(u).\end{multline}
The dissipated heat on the other hand can be written as the sum of the delivered amount of work and the heat dissipation when the temperature of the heat bath is changed:
\begin{multline}\left\langle q\right\rangle_{x,u}=\frac{k_B}{2}\left(T_h-T_c\right)\\-k_B T_h \ln x \left(1+\sqrt{\frac{2}{\pi }}u\ln u\right)+O(u),\end{multline}
after which Eq.~(\ref{etahb}) is found.

\section{Zero temperature engine}
The calculation of the cumulant generating function or large deviation function of the efficiency of the zero temperature engine is a non-trivial problem as efficiency is a non-additive quantity. Instead, we shall first turn to the cumulant generating function of the work and heat:
\begin{equation}K(\lambda,\mu)=\ln\left\langle e^{\lambda w+\mu q}\right\rangle =\ln\left[A(\lambda,\mu)+B(\lambda,\mu)\right],\end{equation}
with:
\begin{equation}A(\lambda,\mu)=\frac{1}{1-k_BT\mu}\left(1-\exp\left(2u^2(k_BT\mu-1\right)\right), \end{equation}
\begin{widetext}
\begin{eqnarray}B(\lambda,\mu)&=&\frac{1}{(1-k_BT\mu)^{\frac{3}{2}}}e^{-2(1+k_BT\lambda)u^2}\left(e^{2k_BT\left(2\lambda+\mu\right)u^2}\sqrt{1-k_BT\mu}\right.\nonumber\\& &\left.+\sqrt{2\pi}\lambda k_BT u\exp\left(2\left(1+\frac{(k_BT\lambda)^2}{1-k_BT\mu}\right)u^2\right) \mathrm{erfc}\left(\sqrt{\frac{2}{1-k_BT\mu}}(1-k_BT(\lambda+\mu))u\right)\right).\nonumber\\\end{eqnarray}
\end{widetext}
This can easily be numerically contracted to find the LDF of the efficiency:
\begin{equation}J(\eta)=-\min_{\lambda}K(\lambda,-\eta\lambda).\end{equation}
In particular, we see that in the limit $\eta\rightarrow 1$:
\begin{equation}
J(\eta)\sim \ln(1-\eta).
\end{equation}

\end{document}